\documentclass[conference]{IEEEtran}
\IEEEoverridecommandlockouts
\usepackage{multirow}
\usepackage{cite}
\usepackage{amsmath,amssymb,amsfonts}
\usepackage{algorithmic}
\usepackage{graphicx}
\usepackage{textcomp}
\usepackage{array}
\usepackage{xcolor}
\usepackage{booktabs}
\usepackage{url}
\def\BibTeX{{\rm B\kern-.05em{\sc i\kern-.025em b}\kern-.08em
    T\kern-.1667em\lower.7ex\hbox{E}\kern-.125emX}}
\begin{document}

\title{Battlefield 5G: Dual-PKI and TPM-Based UE Attestation for Tactical 5G Standalone Networks}



\author{
\IEEEauthorblockN{
Al Nahian Bin Emran$^{*}$, Rajendra Paudyal$^{*}$, Rajendra Upadhyay$^{*}$,
Lisa Donnan$^{*}$,\\Arupjyoti Bhuyan$^{\dagger}$, Duminda Wijesekera$^{*}$ 
}
\IEEEauthorblockA{
$^{*}$Mason Innovation Labs, George Mason University, Arlington, VA, USA\\
$^{\dagger}$Idaho National Laboratory, Idaho Falls, ID, USA\\
\{abinemra$^{*}$, rpaudyal$^{*}$, rupadhya$^{*}$, ldonnan$^{*}$, dwijesek$^{*}$\}@gmu.edu,
arupjyoti.bhuyan@inl.gov$^{\dagger}$
}
}



\maketitle

\begin{abstract}
The standardized 5G Authentication and Key Agreement (5G-AKA) authenticates a subscriber credential stored on a  Universal Subscriber Identity Model (USIM) but does not authenticate the physical device that holds that credential or verify its boot state. This gap is significant in tactical 5G deployments, where user equipment may be captured, modified, returned to service, or used with transplanted subscriber credentials. We present Battlefield 5G, a pre-authentication framework for 5G Standalone networks that combines dual X.509 device-certificate checks with Trusted Platform Module (TPM) -based boot attestation before standard registration is accepted. The design places an outer certificate challenge on the 5G base-station called gNB, an independent inner certificate challenge on the Access and Mobility Management Function (AMF) in the 5G core network, and a TPM PCR (Platform Configuration Register) quote verified by an attestation proxy on the 5G core network side. A gNodeB (gNB) side Radio Resource Control (RRC) forwarding gate and an AMF-side save-and-replay mechanism enable multi-round certificate and attestation challenge-response exchanges to be inserted into the registration path without modifying any 3GPP Non-Access Stratum (NAS) message structures or adding new NAS message types. We implement these capabilities by extending the Radio Access Network of the Software Radio System (srsRAN), gNB, User Equipment of the Software Radio System (srsUE) and Open5GS in a B210-based Universal Radio Peripheral (USRP) testbed with a hardware TPM 2.0 in the UE. The prototype blocks SIM-transplant, rogue-certificate, firmware-tampering, and replay attacks. Across six trials, Battlefield 5G increases average onboarding latency from 1886 ms to 2260 ms, adding 373.4 ms of pre-authentication overhead while preserving standard 5G-AKA, security mode, and packet data unit (PDU) session procedures.
\end{abstract}

\begin{IEEEkeywords}
 5G security, tactical networks, TPM attestation, device authentication, 5G-AKA, RRC, NAS, PKI, Zero Trust
\end{IEEEkeywords}

\section{Introduction}
\label{sec:introduction}

\begin{figure}[!t]
\centering
\includegraphics[width=\columnwidth]{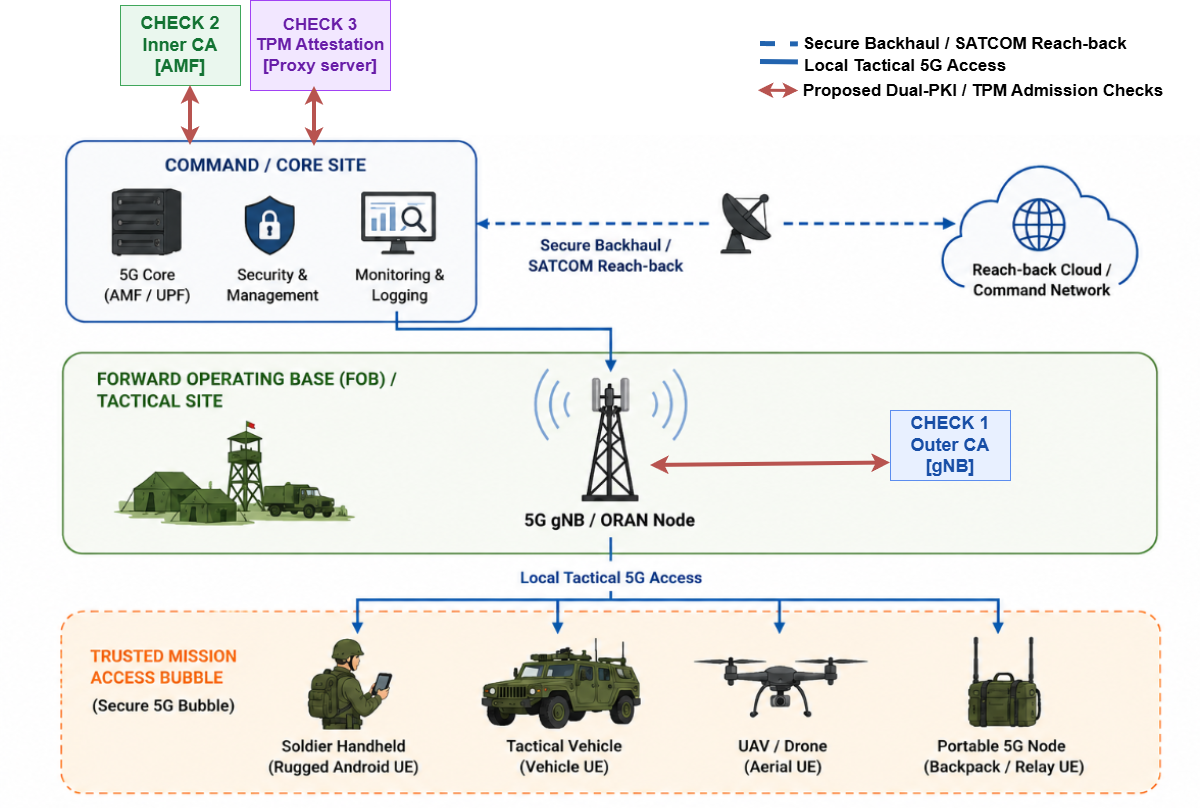}
\caption{Tactical 5G with Battlefield 5G Admission Checks}
\label{fig:tactical_environment}
\end{figure}

The 3GPP 5G security architecture defines 5G-AKA as the primary mechanism to authenticate a subscriber to the network and to obtain keys to protect NAS and access-stratum signaling~\cite{3gpp33501}. However, 5G-AKA authenticates the subscriber credential encoded in the USIM rather than the physical device that holds that credential. As a result, the network learns that a valid subscription is present but does not learn whether the device is enrolled or whether it boots into a trusted software state. This distinction is critical in tactical and battlefield deployments. Because in a forward operating base (FOB), portable/ external private 5G network~\cite{jma2025expeditionary5g,paudyal}, or vehicle-mounted tactical system, the gNB and core can be controlled by the operator, while UEs operate in exposed environments. Furthermore,  a soldier-carried UE, drone controller, vehicle terminal, or field sensor can be captured, modified, or returned to operation. In this operating environment, authenticating only the USIM is insufficient because the network must determine both the validity of the subscriber and the trust of the device. Fig.~\ref{fig:tactical_environment} illustrates the representative tactical 5G environment considered in this work. A command/core site provides 5G core functions, security services, monitoring, and reach-back connectivity, while a forward operating base or tactical site hosts the local 5G gNB/ORAN node. Mission spporting UEs such as soldier handhelds, vehicle terminals, UAVs, and portable 5G nodes access the network through local tactical 5G links. The red arrows in Fig.~\ref{fig:tactical_environment} show the proposed Battlefield 5G admission checks proposed in this paper:  Namely, an Outer CA check in the gNB, an Inner CA check in the AMF, and TPM attestation in the attestation proxy.

Three attacks motivate our solution. First, in a SIM-transplant attack, an adversary moves a valid USIM into an unauthorized platform. Second, in a firmware-tamper attack, a captured UE is modified while retaining valid credentials. Third, in a rogue-device attack, an adversary presents cloned, stolen, or self-issued credentials to the tactical network. 

Our Battlefield 5G addresses this gap by inserting device identity and boot-integrity checks before standard 5G registration is accepted. The dual-certificate design is inspired by the National Security Agency (NSA) Commercial Solutions for Classified (CSfC) key-management guidance, which uses separate Outer and Inner Certificate Authorities (CA) so that the same certificate cannot authenticate between both protection layers~\cite{nsa_csfc_km}. Battlefield 5G adapts this NSA recommended separation principle  by placing an outer X.509 challenge at the gNB over the  Radio Resource Control (RRC) path and an independent inner X.509 challenge at the AMF over the Non-Access Stratum (NAS)/Next Generation Application Protocol (NGAP) path. It then adds a TPM quote over Platform Configuration Registers (PCR) 0--7, verified by an attestation proxy on the core side, to check UE boot integrity. Only after all three checks are successful, AMF replays the saved Registration Request and continues with normal 5G-AKA. The contributions of this paper are as follows:

\begin{itemize}
    \item Adding a dual-rooted PKI pre-authentication protocol that places independent UE certificate checks at the gNB/RRC and AMF/NAS layers before network admission.

    \item Integrating TPM PCR-based boot attestation into the 5G registration path to ensure device integrity.

    \item Implement a gNB-side RRC forwarding gate and an AMF-side save-and-replay mechanism that pauses registration progress, performs certificate and TPM checks, and resumes standard 5G-AKA only after all checks pass.
    
    \item Evaluated the prototype on a USRP-based 5G SA testbed, showing that it blocks four attack scenarios and adds 373.4 ms of pre-authentication overhead over a same-testbed baseline.
\end{itemize}

\section{Background and Related Work}
\label{sec:relatedWork}

\subsection{5G-AKA and Device-Trust Gap}
\label{ssec:5G-AKA}

In the 5G Authentication and Key Agreement (5G-AKA), the Access and Mobility Management Function (AMF) obtains authentication material using the Authentication Server Function (AUSF) and Unified Data Management (UDM), and the Universal Subscriber Identity Module (USIM) computes a response using the long-term subscriber key. Successful authentication establishes a key hierarchy used to derive keys for subsequent Non-Access Stratum (NAS) security between user equipment (UE) and AMF, and Access Stratum (AS) security between UE and next-generation NodeB (gNB)~\cite{3gpp33501}. This authenticates primarily the subscriber credential anchored in the USIM, not the integrity of the physical device or platform hosting that credential. Formal analyzes of 5G-AKA focus on subscriber authentication, key agreement, and privacy properties rather than physical device integrity~\cite{basin2018formal}. Therefore, a valid USIM on a compromised or unauthorized platform can still satisfy the standard access procedure.

\subsection{TPM-Based Remote Attestation}
\label{ssec:tpm}

A TPM 2.0 provides hardware-protected keys and Platform Configuration Registers (PCRs) for measured boot~\cite{tcg2020tpm}. During boot, the firmware and software components are measured and extended to PCRs that use $PCR_i^{(t+1)}=H(PCR_i^{(t)}\parallel m_t)$,
where $m_t$ is the measurement at time $t$. A verifier first sends a fresh nonce to attester to prevent replay, and attester uses the UE's TPM to generate a quote over the selected PCR values and the verifier-supplied nonce, and signs the quote using the TPM-resident Attestation Key (AK). The verifier then checks the AK public key or certificate, verifies the quote signature, compares the reported PCR values with the reference values, and produces an attestation result. The Internet Engineering Task Force (IETF) Remote ATtestation ProcedureS (RATS) architecture defines this general pattern using attesters, verifiers, evidence, reference values, and attestation results~\cite{birkholz2023rats}. In Battlefield 5G, the UE is the attester, the proxy is the verifier, and the AMF uses the attestation result as an admission decision.

\subsection{Cellular Bootstrapping and Tactical 5G}
\label{ssec:bootstrap}

Hussain et al. showed that insecure connection bootstrapping is a root cause of cellular attacks~\cite{hussain2019bootstrapping}. Singla et al. proposed secure 5G bootstrapping to defend against fake base stations~\cite{singla2021look}. Other work explored public-key protection for system information and base-station authentication~\cite{ross2024fixing,purification2025base,dong2025securing,lotto2023baron}. These works authenticate the network to the UE. In contrast, Battlefield 5G authenticates the UE device to the network before registration is accepted. Bhuyan et al. demonstrated secure 5G roaming between private networks for military communications, using Open5GS as the 5G core in test networks operated by Idaho National Laboratory (INL) and the Cooperative Cyber Defense Center of Excellence (CCDCOE)~\cite{bhuyan2024real}. Their work shows that open-source 5G core implementations can support defense-oriented experimentation and real-world military communication scenarios.

Remote attestation has also been used in cloud and edge environments. Keylime provides TPM-based attestation to maintain platform trust~\cite{schear2016bootstrapping}. Remote attestation has also been used in cloud and edge environments. 
Our prior work extended Keylime with TPM 2.0 and IMA to provide continuous pod-level integrity verification for Kubernetes-deployed 5G VNFs such as AMF, SMF and UPF~\cite{11410158}. In contrast, this work moves the attestation into the User Equipment (UE) admission path, requiring platform evidence before the device receives normal network access. In addition, Zero Trust architectures require access decisions to consider identity and device posture rather than implicit network location~\cite{dod2022zta}. Battlefield 5G applies this principle to tactical 5G by requiring device certificates and TPM evidence before the standard 5G registration path continues.

\section{The Threat Model}
\label{sec:threatModel}

We consider an adversary that can capture a legitimate UE, extract or reuse its USIM, modify firmware or boot components, deploy an SDR-based rogue platform, observe over-the-air messages, and replay previously captured certificate or attestation responses. We assume that this adversary cannot extract TPM-resident private keys, cannot forge signatures from uncompromised certificate authority (CA) keys, and cannot make modified boot measurements match a known-good PCR baseline without booting the expected software stack. The gNB, AMF, and attestation proxy are trusted components in this prototype.

Table~\ref{tab:threats} maps the main attack classes to the corresponding protection mechanisms in Battlefield 5G. The system enforces four goals before registration is accepted: (1) device identity, where the UE must prove possession of an enrolled device private key; (2) checkpoint separation, where compromise of one certificate checkpoint does not bypass all admission checks; (3) boot integrity, where TPM PCRs must match an enrolled baseline; and (4) replay resistance, where captured responses cannot satisfy fresh challenges.

\begin{table}[!t]
\caption{Threats and Protection Mechanisms}
\label{tab:threats}
\centering
\footnotesize
\begin{tabular}{p{0.42\columnwidth}p{0.48\columnwidth}}
\toprule
\textbf{Threat} & \textbf{Protection Mechanism} \\
\midrule
SIM transplant & Outer X.509 check at gNB before NGAP forwarding \\
Rogue certificate & Certificate-chain validation against trusted CA bundle \\
Firmware or boot tampering & TPM quote over PCRs 0--7 compared with baseline \\
Replay of captured evidence & Fresh nonce bound into certificate response and TPM quote \\
Single checkpoint compromise & Independent outer and inner CA hierarchies \\
\bottomrule
\end{tabular}
\end{table}

\section{System Design}
\label{sec;design}

\subsection{Overview}
\label{ssec:designOverview}

\begin{figure}[!t]
\centering
\includegraphics[width=\columnwidth]{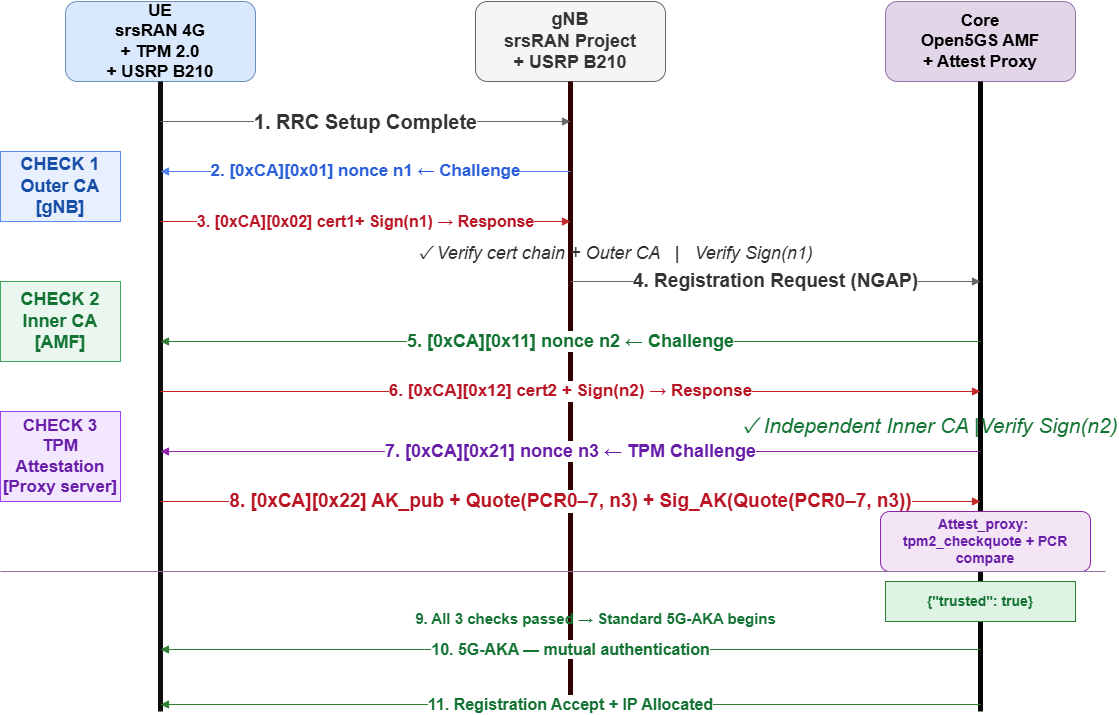}
\caption{Triple-check pre-authentication flow across the UE, gNB, and core network.}
\label{fig:triple_check_flow}
\end{figure}

Battlefield 5G inserts three pre-authentication checks before standard 5G-AKA processing, as shown in Fig.~\ref{fig:triple_check_flow}. Check~1 is an outer device-certificate challenge issued by the gNB after RRC setup and before NAS forwarding (Steps~1--3); if this check succeeds, the gNB forwards the Registration Request to the AMF (Step~4). Check~2 is an independent inner device-certificate challenge issued by the AMF before the saved Registration Request is processed (Steps~5--6). Check~3 is a challenge to the TPM attestation issued by the AMF and verified by an attestation proxy (Steps~7--8). Only after passing all three checks does the AMF replay the saved Registration Request and continue standard 5G registration, where 5G-AKA and registration completion occur (Steps~9--11).

\subsection{Dual-Root PKI and Wire Protocol}
\label{ssec:DualRoot+WireProtocol}

The system uses two independent certificate hierarchies. The outer hierarchy is trusted only by the gNB, while the inner hierarchy is trusted only by the AMF:
\begin{align}
\text{Outer Root CA} &\rightarrow \text{Outer CA} \rightarrow \text{UE Outer Cert},\\
\text{Inner Root CA} &\rightarrow \text{Inner CA} \rightarrow \text{UE Inner Cert}.
\end{align}

The UE stores both certificates and private keys. The gNB stores only the outer CA bundle and the AMF stores only the inner CA bundle. Thus, the same credential cannot satisfy both checkpoints. A UE that passes the gNB checkpoint must still prove possession of an unrelated inner certificate before the AMF accepts registration.

All pre-authentication messages are encoded using a compact binary wire protocol that begins with a magic byte \texttt{0xCA} followed by a message type. The messages are transported over existing RRC DL/UL Information Transfer paths, avoiding new RRC message types. Certificate responses sign a fresh 32-byte nonce using the corresponding UE private key. The verifier validates the certificate chain and verifies the nonce signature using the public key in the certificate. Table~\ref{tab:wire} lists the custom pre-authentication message types used for the outer certificate, inner certificate, and TPM attestation exchanges.

\subsection{RRC Forwarding Gate, Save-and-Replay Mechanism and TPM Attestation Gate}
\label{ssec;RRC}

The proposed protocol uses two control points to insert pre-authentication checks without changing 3GPP NAS message formats. The first control point is at the gNB. After RRC connection setup, the UE's NAS Registration Request is carried through the RRC path toward the core. The gNB does not immediately forward this registration message to the AMF. Instead, it first performs the Outer CA challenge-response exchange over the RRC path. During this exchange, the registration message is preserved by the gNB-side control flow and is forwarded to the AMF only if the outer certificate verification succeeds. If verification fails, the gNB releases the RRC connection, and the AMF never receives the UE's registration attempt.

\begin{table}[!t]
\caption{Custom Pre-Authentication Message Types}
\label{tab:wire}
\centering
\footnotesize
\begin{tabular}{lll}
\toprule
\textbf{Message} & \textbf{Type} & \textbf{Payload} \\
\midrule
OUTER\_CHALLENGE & \texttt{0x01} & 32-byte nonce \\
OUTER\_RESPONSE & \texttt{0x02} & DER cert, signature \\
INNER\_CHALLENGE & \texttt{0x11} & 32-byte nonce \\
INNER\_RESPONSE & \texttt{0x12} & DER cert, signature \\
TPM\_CHALLENGE & \texttt{0x21} & 32-byte nonce \\
TPM\_RESPONSE & \texttt{0x22} & AK public, quote, signature, PCRs \\
\bottomrule
\end{tabular}
\end{table}

The second control point is at the AMF, where both the Inner CA check and the attestation TPM check are enforced. Open5GS normally processes a Registration Request immediately through the 5G Mobility Management (GMM) handler. To insert these checks, the AMF saves the raw NAS packet buffer and associated event state in the UE context, then returns before normal registration processing. The AMF first sends the Inner CA challenge. After the inner certificate response is verified, the saved Registration Request is re-injected into the AMF event path.

The TPM attestation gate then intercepts the replayed request before normal registration continues. During enrollment, each UE is booted in a known-good state, and the verifier records the UE's TPM Attestation Key (AK) identity and baseline PCR values. At connection time, after the inner certificate check succeeds, the AMF sends a fresh TPM challenge nonce to the UE. The UE invokes the hardware TPM to generate a quote for PCRs 0--7 and returns the AK public key, TPM quote, quote signature, and PCR values using the \texttt{TPM\_RESPONSE} message. PCRs 0--7 are used because they represent early measured-boot state, including firmware, bootloader, and early operating-system measurements. The AMF forwards this evidence to the attestation proxy, which verifies that the AK matches the enrolled UE, checks the quote signature, confirms that the nonce matches the current challenge, and compares the reported PCR values against the enrolled baseline. If all checks pass, the proxy returns a trusted result and the AMF replays the saved Registration Request. If any check fails, the AMF rejects the registration attempt.

Together, the gNB-side RRC forwarding gate, the AMF-side save-and-replay mechanism, and the TPM attestation gate provide a suspension-and-resumption path for pre-registration security checks. The RRC forwarding gate enforces the Outer CA check before the UE's registration message reaches the core, while the AMF save-and-replay mechanism enables the Inner CA and TPM attestation checks before normal registration processing continues. This allows Battlefield 5G to enforce all three checks before standard 5G-AKA proceeds, while leaving the 3GPP NAS message format unchanged.

\section{Implementation}
\label{sec:implementation}

The prototype changes srsRAN Project \cite{srsran}, srsRAN 4G/srsUE, and Open5GS \cite{open5gs}, and adds a Python Flask attestation proxy. The proxy verifies the evidence and compares the PCR values compared to the enrollment baseline. At the gNB, the outer challenge is generated after RRC setup and before the UE's registration message is forwarded to the core. The gNB-side RRC forwarding path preserves the Registration Request while the Outer CA challenge-response exchange is performed. If the UE provides a valid outer certificate response, the original Registration Request is forwarded to the AMF through NGAP. If verification fails, the RRC connection is released and no NGAP traffic reaches the AMF. At the AMF, the GMM Registration Request path is modified to save and replay the Registration Request while the Inner CA and TPM gates are evaluated before normal registration processing. The AMF checks the \texttt{0xCA} magic byte and routes pre-authentication responses to the certificate-authentication and attestation logic instead of normal NAS processing. The proxy exposes an HTTP endpoint that receives base64-encoded TPM evidence, verifies the quote, checks the AK hash and PCR values, and returns a JSON trust result.

\begin{table*}[!t]
\centering
\caption{Attack Scenario Verification Across the Dual-PKI and TPM Attestation Pipeline}
\label{tab:attack_verification}
\footnotesize
\setlength{\tabcolsep}{4pt}
\renewcommand{\arraystretch}{1.12}
\begin{tabular}{
>{\raggedright\arraybackslash}p{0.21\textwidth}
>{\centering\arraybackslash}p{0.15\textwidth}
>{\centering\arraybackslash}p{0.15\textwidth}
>{\centering\arraybackslash}p{0.18\textwidth}
>{\raggedright\arraybackslash}p{0.25\textwidth}
}
\toprule
\textbf{Attack Scenario} &
\textbf{Outer CA Check (gNB/RRC)} &
\textbf{Inner CA Check (AMF/NAS)} &
\textbf{TPM Attestation Check (AMF + Proxy)} &
\textbf{Detection Point and Outcome} \\
\midrule

SIM transplant &
\textbf{Fail} &
Not reached &
Not reached &
Blocked at the gNB because the UE lacks the required outer device certificate/private key. \\

Rogue certificate &
\textbf{Fail} &
Not reached &
Not reached &
Blocked at the gNB because the certificate is self-signed or not issued by the trusted Outer CA. \\

Firmware tamper simulation &
Pass &
Pass &
\textbf{Fail} &
Blocked after TPM verification because the quoted PCR values do not match the enrolled baseline. \\

Replay attack &
\textbf{Fail if replayed} &
\textbf{Fail if replayed} &
\textbf{Fail if replayed} &
Blocked by nonce freshness at the replayed checkpoint; certificate responses and TPM quotes cannot be reused across sessions. \\

\bottomrule
\end{tabular}
\end{table*}

\section{Evaluation}
\label{sec:evaluation}

\subsection{Experimental Setup}
\label{ssec:setup}

The system was evaluated on a three-node 5G SA testbed. The UE and gNB used USRP B210 radios. The core ran Open5GS and the attestation proxy in a Docker network. The UE used a TPM 2.0 device and \texttt{tpm2-tools} for quote generation. Each latency value in Tables~\ref{tab:baseline_battlefield_comparison} and~\ref{tab:preauth_latency_breakdown} is derived from six successful UE onboarding trials.

\subsection{Functional and Attack Validation}

A successful connection requires that all three checks pass. First, the gNB verifies the outer certificate chain and the nonce signature. Second, the AMF verifies the independent inner certificate. Third, the AMF obtains a trusted result from the attestation proxy after verifying the TPM quote. Only after these checks pass does the AMF process the saved Registration Request and continue normal 5G-AKA.

Table~\ref{tab:attack_verification} summarizes where each attack is stopped. Credential-only attacks are blocked at the gNB before reaching the core, while attacks involving valid certificates but invalid platform state are blocked by TPM attestation at the core side.

\subsection{Latency, Overhead and Performance}
\label{ssec:latency}

To quantify end-to-end impact, we evaluated two configurations on the same physical testbed. The baseline configuration is an unmodified 5G SA using Open5GS, srsRAN Project for gNB, and srs4G for UE without certificate authentication or TPM attestation. The Battlefield 5G configuration enables the proposed Outer CA, Inner CA, and TPM attestation checks.  Table~\ref{tab:baseline_battlefield_comparison} reports the end-to-end onboarding latency for both configurations, while Table~\ref{tab:preauth_latency_breakdown} breaks down the added pre-authentication latency by checkpoint.

\begin{table*}[!t]
\centering
\caption{Baseline 5G and Battlefield 5G Onboarding Latency Comparison}
\label{tab:baseline_battlefield_comparison}
\scriptsize
\setlength{\tabcolsep}{3pt}
\renewcommand{\arraystretch}{1.12}
\begin{tabular}{
>{\raggedright\arraybackslash}p{0.25\textwidth}
>{\raggedleft\arraybackslash}p{0.10\textwidth}
>{\raggedleft\arraybackslash}p{0.08\textwidth}
>{\raggedleft\arraybackslash}p{0.08\textwidth}
>{\raggedleft\arraybackslash}p{0.11\textwidth}
>{\raggedleft\arraybackslash}p{0.08\textwidth}
>{\raggedleft\arraybackslash}p{0.08\textwidth}
}
\toprule
\multirow{2}{*}{\textbf{Phase}} &
\multicolumn{3}{c}{\textbf{Baseline 5G}} &
\multicolumn{3}{c}{\textbf{Battlefield 5G}} \\
\cmidrule(lr){2-4}
\cmidrule(lr){5-7}
&
\textbf{Mean} &
\textbf{Min} &
\textbf{Max} &
\textbf{Mean} &
\textbf{Min} &
\textbf{Max} \\
\midrule

gNB RRC setup &
1448 ms &
1422 ms &
1475 ms &
1452 ms &
1426 ms &
1479 ms \\

Outer CA -- Check 1 &
-- &
-- &
-- &
74.9 ms &
47.0 ms &
89.0 ms \\

Inner CA + TPM -- Check 2+3 &
-- &
-- &
-- &
298.5 ms &
274.0 ms &
308.0 ms \\

Standard 5G (AKA + NAS + PDU) &
438 ms &
419 ms &
459 ms &
435 ms &
417 ms &
456 ms \\

Total onboarding latency &
1886 ms &
1841 ms &
1934 ms &
2260 ms &
2228 ms &
2303 ms \\

Cert-auth and attestation overhead &
-- &
-- &
-- &
373.4 ms &
361.9 ms &
386.0 ms \\

\bottomrule
\end{tabular}
\end{table*}

Table~\ref{tab:baseline_battlefield_comparison} shows that the baseline configuration averages 1886 ms total onboarding latency, while Battlefield 5G averages 2260 ms. The difference corresponds to 373.4 ms of added pre-authentication overhead. This value is not the total time-to-IP; rather, it is the additional delay introduced by the certificate-authentication and TPM-attestation layer before and during the standard 5G onboarding path.

\begin{table*}[!t]
\centering
\caption{Pre-authentication Latency Breakdown}
\label{tab:preauth_latency_breakdown}
\scriptsize
\setlength{\tabcolsep}{2pt}
\renewcommand{\arraystretch}{1.12}
\begin{tabular}{
>{\raggedright\arraybackslash}p{0.145\textwidth}
>{\raggedright\arraybackslash}p{0.105\textwidth}
>{\raggedright\arraybackslash}p{0.285\textwidth}
>{\raggedleft\arraybackslash}p{0.065\textwidth}
>{\raggedleft\arraybackslash}p{0.085\textwidth}
>{\raggedright\arraybackslash}p{0.170\textwidth}
}
\toprule
\textbf{Phase} &
\textbf{Scope} &
\textbf{What is Included} &
\textbf{Mean} &
\textbf{Min--Max} &
\textbf{Interpretation} \\
\midrule

\textbf{Outer CA certificate RTT} &
gNB $\rightarrow$ UE $\rightarrow$ gNB &
RRC downlink challenge, UE response generation/signing, RRC uplink response, and gNB verification completion &
\textbf{74.9 ms} &
47.0--89.0 ms &
Full outer certificate challenge-response delay over the radio interface. \\

\textbf{Outer CA OpenSSL verification} &
Local gNB only &
DER parsing, X.509 chain verification, and RSA-SHA256 signature verification &
\textbf{1.0 ms} &
0.9--1.2 ms &
Pure cryptographic verification cost at the gNB. \\

\textbf{Inner CA certificate RTT} &
AMF $\rightarrow$ gNB $\rightarrow$ UE $\rightarrow$ gNB $\rightarrow$ AMF &
NAS/NGAP challenge delivery, RRC air-interface transit, UE-side RSA signing, return path, and AMF certificate validation &
\textbf{49.2 ms} &
40.0--54.0 ms &
Full inner certificate challenge-response delay, not only cryptographic processing. \\

\textbf{TPM attestation RTT} &
AMF $\rightarrow$ UE $\rightarrow$ AMF &
TPM challenge delivery, UE-side \texttt{tpm2\_quote}/\texttt{tpm2\_readpublic}, quote packaging, and response return to AMF &
\textbf{235.7 ms} &
219.0--258.0 ms &
Dominant delay; this is challenge-response RTT, not pure TPM hardware time. \\

\textbf{Attestation proxy HTTP RTT} &
AMF $\rightarrow$ proxy $\rightarrow$ AMF &
HTTP POST from AMF to proxy, \texttt{tpm2\_checkquote}, PCR comparison, and JSON response &
\textbf{13.6 ms} &
11.0--16.0 ms &
Includes proxy processing and Docker/container network round trip. \\

\textbf{Total AMF-side pre-authentication latency} &
AMF-side gate delay &
Inner CA RTT + TPM attestation RTT + proxy verification path until registration replay &
\textbf{298.5 ms} &
274.0--308.0 ms &
Added AMF-side delay before normal 5G registration proceeds. \\

\textbf{Estimated total pre-authentication overhead} &
gNB + AMF path &
Outer CA certificate RTT + AMF-side pre-authentication latency &
\textbf{373.4 ms} &
361.9--386.0 ms &
Full added delay from dual-PKI and TPM attestation. \\

\bottomrule
\end{tabular}
\end{table*}

Table~\ref{tab:preauth_latency_breakdown} separates radio/signaling RTT from local cryptographic processing. The Outer CA exchange averages 74.9 ms, but the local OpenSSL verification at the gNB takes only 1.0 ms, showing that certificate processing is not the bottleneck. The Inner CA exchange averages 49.2 ms and includes NAS/NGAP forwarding, RRC transit, UE-side signing, and AMF-side verification. The RTT TPM attestation dominates the added delay at 235.7 ms, or approximately 79\% of the 298.5 ms AMF-side pre-authentication latency. This cost is primarily due to the generation of TPM quotes on the UE-side, the retrieval of AK public-keys, and the construction of response. The system also adds 3609 bytes of signaling per connection: 1100 bytes for the Outer CA exchange, 1100 bytes for the Inner CA exchange, and 1409 bytes for the TPM attestation exchange. This overhead is incurred once per UE onboarding attempt.

As seen,  certificate verification is lightweight (~1ms), but TPM attestation dominates the added delay (235.7ms). Consequently, our ongoing work addresses the reduction of TPM latency command by replacing command-line TPM tools with direct TSS library calls or securely caching attestation artifacts where appropriate. Compared with unmodified 5G SA, Battlefield 5G increases the average total onboarding latency from 1886 ms to 2260 ms. This corresponds to less than 0.4 seconds of added pre-authentication overhead. 

\section{Security Analysis}
\label{sec:security}

The proposed pipeline enforces three complementary properties before network admission: device identity in the gNB, independent device identity in the AMF, and boot integrity  in the attestation proxy. SIM-transplant and rogue-certificate attacks fail at the Outer CA Check because the attacker cannot produce a valid certificate chain and nonce signature under the trusted outer CA. Thus, unauthorized devices are blocked before NAS traffic reaches the core (Table~\ref{tab:attack_verification}). The dual-CA design provides defense in depth. Passing the gNB checkpoint does not imply acceptance by the AMF, because the AMF verifies a separate inner certificate under an independent CA hierarchy.   Therefore, both a compromise and a misconfiguration of one  check point will not automatically bypass the other. TPM attestation prevents valid-but-compromised devices from entering the network. Firmware tampering changes PCR values, causing the proxy to reject the quote even when both certificates are valid. Replay is prevented because all certificate responses and TPM quotes are bound to fresh nonces; captured evidence fails at whichever checkpoint is replayed. The UE is allowed to proceed to standard 5G-AKA only after all three checks are successful.

The current prototype verifies PCRs 0--7, which cover early boot integrity but not runtime user-space changes. Runtime integrity can be added using PCR 10 and IMA event-log verification. In addition, the current enrollment database is a flat file. Verifying runtime integrity by using PCR-10 checks and replacing the flat file with a scalable database for larger troop deployments are our immediate ongoing work.


\section{Conclusion}
\label{sec:conclusions}

We present Battlefield 5G, a pre-authentication framework for tactical 5G SA networks. Our system addresses a gap in standard 5G-AKA by verifying the UE device and its boot state before registration is accepted. To do so, we combine two independent certificates, an outer X.509 certificate check at the gNB, an independent inner X.509 certificate check at the AMF, and a TPM PCR-based attestation verified by an attestation proxy. A gNB-side RRC forwarding gate and an AMF-side save-and-replay mechanism enable these checks to be inserted into the registration path without changing 3GPP NAS message structures. The prototype blocks SIM-transplant, rogue-certificate, firmware-tamper, and replay attacks, while adding 373.4 ms of pre-authentication overhead over a same-testbed baseline. These results show that certificate-based device identity and hardware-rooted boot integrity verification can be made a part of the 5G access decision for tactical networks. In contested environments where compromised or unauthorized devices must not be admitted to the network, this pre-admission layer provides a practical and deployable mechanism on commodity 5G hardware without changing 3GPP NAS message structures or replacing standard 5G-AKA, security mode, and PDU session procedures.

\bibliography{custom}
\bibliographystyle{IEEEtran}

\end{document}